\documentclass[%
 aps,
 prl,%
 amsmath,amssymb,
preprint,%
]{revtex4-1}

\usepackage{graphicx}
\usepackage{dcolumn}
\usepackage{bm,mathdots}

\begin{document}

\preprint{}

\title[Sample title]{Double Dressing and Manipulation of the Photonic \\ Density of States in Nanostructured Qubits}

\author{Hanz Y. Ram\'irez}
 \email{hanzyrg@hotmail.com}
\affiliation{ 
Hefei National Laboratory for Physics at the Microscale, University of Science and Technology of China, Hefei, Anhui 230026, China 
}%


\date{\today}

\begin{abstract}
In this work, a model to study the coupling between a semiconductor qubit and two time-dependent electric fields is developed. By using it in the resonantly monochromatic double dressing regime, control of the local density of optical states is theoretically and numerically demonstrated for a strongly confined exciton. 

Drastic changes in the allowed energy transitions yielding tunable broadening of the optically active frequency ranges, are observed in the simulated emission spectra. The presented results are in excellent qualitative and quantitative agreement with recent experimental observations. 
\end{abstract}

\maketitle

%

\section{1. Introduction}

Few decades ago, artificial atoms emerged timidly in the realm of quantum optics as potential photon sources, eventually alternative to their well established natural counterpart \cite{quantum-dot}. Nowadays, due to progress in material fabrication and characterization techniques, they have become not just plausible, but in cases even more suitable for both, basic research and technological applications. One outstanding example is the initially unambiguous and now impressively diaphanous observation of dressed states in self assembled
quantum dots \cite{nat-MT-1,nat-MT-2,nat-nano-ustc}. This makes nanostructured qubits (NQ) most attractive systems for doubly dressing endeavors \cite{atomicsimilar1,atomicsimilar2,atomicsimilar3,parallel}.

On the other hand, key properties of optoelectronic systems and devices such as emission [absorption] frequencies and intensities, are directly determined by the local density of optical states (LDOS); which basically describes how efficient a system is to emit [absorb] photons at some particular energy. In those devices, the LDOS is set by the electronic density of energy states, so that reduction in the dimensionality rises sharpness in the ranges of optically accessible energies. These changes achieve a limit when all the spacial dimensions
are comparable to the wavelength associated to the confined charge carriers (0D systems), case in which the LDOS distribution consists of Dirac deltas (actually narrow Lorentzians, as in high quality NQs) \cite{subnormal-prl}.

In this work, a size-independent method to control the LDOS of a 0D semiconductor emitter is described; thus providing an indeed realizable scheme to generate photonic subbands of tunable width from fully discretized states.

The paper is organized as follows; first a general formulation to describe slightly detuned double driven NQs and simulate their resonance fluorescence spectra, is presented. Then, special attention is paid to the monochromatic double dressing case in which photonic subband generation is demonstrated. In the last part an analytical expression is derived for the relevant case under study, and conclusions are drawn.

\section{2. Theoretical description}

The NQ is modeled as a semiconductor quantum dot (QD) with well defined vacuum and S-type exciton states, separated by energy $\hbar\omega_0$, under continuous stimulus by two lasers of frequencies $\omega_A$ and $\omega_B$, respectively [see figure 1(a)]. 

The total Hamiltonian for the two distinguishable fields interacting with the exciton in the dot, in the Jaynes-Cummings framework reads 

\begin{eqnarray}
\label{eq-1}
\hat{H} &=& \frac{1}{2}\hbar \omega_0 \hat{\sigma}_z + \hbar \omega_A (\hat{n}+\frac{1}{2}) + \hbar g (\hat{a} \hat{\sigma}_+ + c.c.) \nonumber \\ &\hspace*{1ex}& + \hbar \omega_B (\hat{m}+\frac{1}{2}) + \hbar g (\hat{b} \hat{\sigma}_+ + c.c.) + E_{off} \hspace*{1ex} ,
\end{eqnarray}

where $\hat{\sigma}_+$ is the rising part of the dot dipole transition operator, $g$ is the dot-field coupling, and $\hat{n}$ and $\hat{a}$ ($\hat{m}$ and $\hat{b}$) are the number of photons and  photon annihilation operators for the laser A (B), respectively \cite{jaynescummings,qopticsbook}. $ E_{off} $ is an offset than allows the energy reference to be chosen at convenience.

Currently, highly confined quantum dots exhibit typical neutral exciton excitation energy at the eV scale \cite{prb-hanz}, so that detunings $\delta_A\equiv\omega_0 - \omega_A$ and $\delta_B\equiv\omega_0 - \omega_B$ at the order of terahertz or even far infrared, can be considered much smaller than the resonance frequency $\omega_0$.    

Let us assume $\omega_0 \geq \omega_A \geq \omega_B $ and take as basis the triple direct product between the energy eigenstates of the dot and each of the fields, i.e. $\mid k , n , m \rangle$ where $k=1,2$ and $n,m=0,1,2,3...$ It can be noted that if the coupling $g$ is artificially turned off, for every fixed integer $K>0$ all states with $k=0$ and $n+m=K$, alongside of all states with $k=1$ and $n+m=K-1$; generate a cluster with $2K+1$ eigenenergies ranging from $E_1 = K \hbar \omega_B$ (corresponding to $\mid 0, 0 , K \rangle$) to $E_{2K+1}= (K-1) \hbar \omega_A + \hbar \omega_0$ (corresponding to $\mid 1, K-1 , 0 \rangle$).


Now, if the coupling effects are taken into account, the off diagonal non zero terms are  

\begin{eqnarray}
\langle m,n,1 \mid \hat{H}^C \mid 0,n+1 , m \rangle &=& \hbar g \sqrt{n+1} \hspace*{1ex} , \nonumber \\ 
 \langle m,n,1 \mid \hat{H}^C \mid 0, n , m+1 \rangle &=&  \hbar g \sqrt{m+1} \hspace*{1ex}, 
\end{eqnarray}

and their corresponding complex conjugates. 

For a given $K$, we can rewrite the basis elements in the form  $\mid k , K-m , m \rangle$, which emphasizes the number of photons in the laser B as compared to the total of photons. We introduce the label $l$ for the elements of the basis in the corresponding cluster; i.e. $l=1,2,3,...,2K,2K+1$. Thus we can notate by $\mid \phi_l^K \rangle$ the basis of a determined subspace with total number of photons $K$ (see inset in Fig. 1). The index $l$ is related to the number of photons in the laser B and to the QD state through

\begin{eqnarray}
\label{eq-3}
k=0 (k=1) \hspace*{1ex} \wedge \hspace*{1ex} m = \frac{l-1(l-2)}{2} \hspace*{1ex} ; \hspace*{1ex} \textrm{If} \hspace*{1ex} l \hspace*{1ex} \textrm{odd (even)}   \hspace*{1ex} .
\end{eqnarray} 

Once the Hamiltonian in eq. (1) is written in the basis $\mid \phi_l^K \rangle$, it can be numerically diagonalized to obtain $2K+1$ eigenvalues with their corresponding eigenvectors. We represent with $\mid \psi_l^K \rangle$ the orthonormal basis formed by the eigenstates of the coupled system; that is 

\begin{equation}
\label{eq-4}
\mid \psi_l^K \rangle=\sum_{j=1}^{2K+1} c_{j,l}^K \mid \phi_j^K \rangle \hspace*{1ex} .   
\end{equation}

The coefficients $c_{j,l}^K = \langle \phi_j^K \mid \psi_l^K \rangle$, are directly obtained from the columns of the unitarian matrix $\hat{P}$ that diagonalizes the Hamiltonian, in such a way that $\hat{P}^{-1}\hat{H}\hat{P}=\hat{H}'$ (where $\hat{H}'$ is diagonal). They evidently represent the projection of the states of the uncoupled basis on the eigenvectors of the coupled system.

When two lasers with specified powers are applied on the dot, the numbers of photons in lasers A and B interacting with the dot states are set; let us say $N$ and $M$, respectively ($N+M=K$). This means that the optically accessible states of the coupled system are those in which the specific state $\mid 0 , K-M , M \rangle \equiv \mid \phi_{2M+1}^K \rangle $, is part of the superposition. Hence, under such conditions the LDOS of the coupled system associated to the reference state $\mid \phi_{2M+1}^{K} \rangle^\dag$, is  

\begin{equation}
\label{eq-5}
\rho_{M}^K (E)=\sum_l \mid \langle \phi_{2M+1}^{K} \mid \psi_l^{K} \rangle \mid^2 \delta(E - E_{l}^K) \hspace*{1ex}.
\end{equation}

We are interested in the strong coupling regime, where the Rabi splitting as compared to the QD emission linewidth is large enough to allow steady exciton population. In this case the inelastic part of the light-matter scattering is the dominant one \cite{qopticsbook}, and the emission intensity in a particular frequency $\omega$ will be proportional to the transition rate of the system releasing a photon of energy $\hbar\omega$. If the transformation $P$ is known [and consequently the LDOS $\rho_{M}^{K}(E)$], the fluorescence spectrum can be obtained from the Fermi golden rule (FGR) \cite{natureFGR,scienceFGR,fermigoldenrule}. Namely

\begin{equation}
\label{app-1}
I(\omega) \propto \sum_{F,I} \mid \langle \psi_F^K \mid \hat{\sigma}_- \mid \psi_I^{K+1} \rangle \mid^2 \rho_{M}^{K} (E_F) \delta(\hbar\omega + E_F - E_I) \hspace*{1ex},
\end{equation}

where $\langle \psi_F^K \mid \hat{\sigma}_- \mid \psi_I^{K+1} \rangle$ is the interaction matrix element in which the lowering operator turns the initial state $\mid I \rangle$ into the final state $\mid F \rangle^\dag$, and $\rho_{M}^{K}(E_F)$ stands for the availability of the final state in the $I \rightarrow F$ transition. 

In order to include the finite exciton lifetime for simulating realistic spectra, the Dirac delta distribution in the FGR can be replaced by a Lorentzian distribution $L(\hbar\omega - E_I - E_F ; \Gamma_{I,F})=\frac{1}{\pi}\frac{\hbar \Gamma_{I,F}}{(\hbar\omega - E_I + E_F)^2 + (\hbar \Gamma_{I,F})^2}$, where $\Gamma_{I,F}$
is the $I \rightarrow F$ transition spectral linewidth (Wigner-Weisskopf approximation) \cite{wignerberman,bookdevices}  

For an experimentally set total number of photons ``$K$", the dominant emitting transitions are those from initial states in the cluster corresponding to the subspace $K+1$ to final states in the cluster corresponding to the subspace $K$. 


%
The relevant matrix element are then those of the form

\begin{equation}
\label{eq-7}
\langle \psi_{F}^{K} \mid \hat{\sigma}_- \mid \psi_I^{K+1} \rangle =  \sum_{j_1=1}^{2K+1} \sum_{j_2=1}^{2K+3}   c_{j_1,F}^{K} c_{j_2,I}^{K+1} \langle \phi_{j_1}^K \mid \hat{\sigma_-} \mid \phi_{j_2}^{K+1} \rangle \hspace*{1ex},
\end{equation}

where the ``$*$" in the coefficients of the final states has been omitted because the elements of the transformation $P$ are all real.

Considering the action of the operator $\hat{\sigma_-}$, the correspondences shown in eq. (\ref{eq-3}), and the orthonormality of the basis $\mid \phi_l^K \rangle$; the required dipole transition matrix elements can be expressed in the compact form

\begin{equation}
 \mid \langle \psi_F^{K} \mid \hat{\sigma}_- \mid \psi_I^{K+1} \rangle \mid^2 \hspace*{1ex} = \hspace*{1ex} \mid \sum_{j=2,4,6,...}^{2K+2} c_{j-1,F}^{K} c_{j,I}^{K+1} \mid^2 \hspace*{1ex}.
\end{equation}


\section{3. Monochromatic double dressing}

\begin{figure}[t]
\scalebox{1.0}{\includegraphics{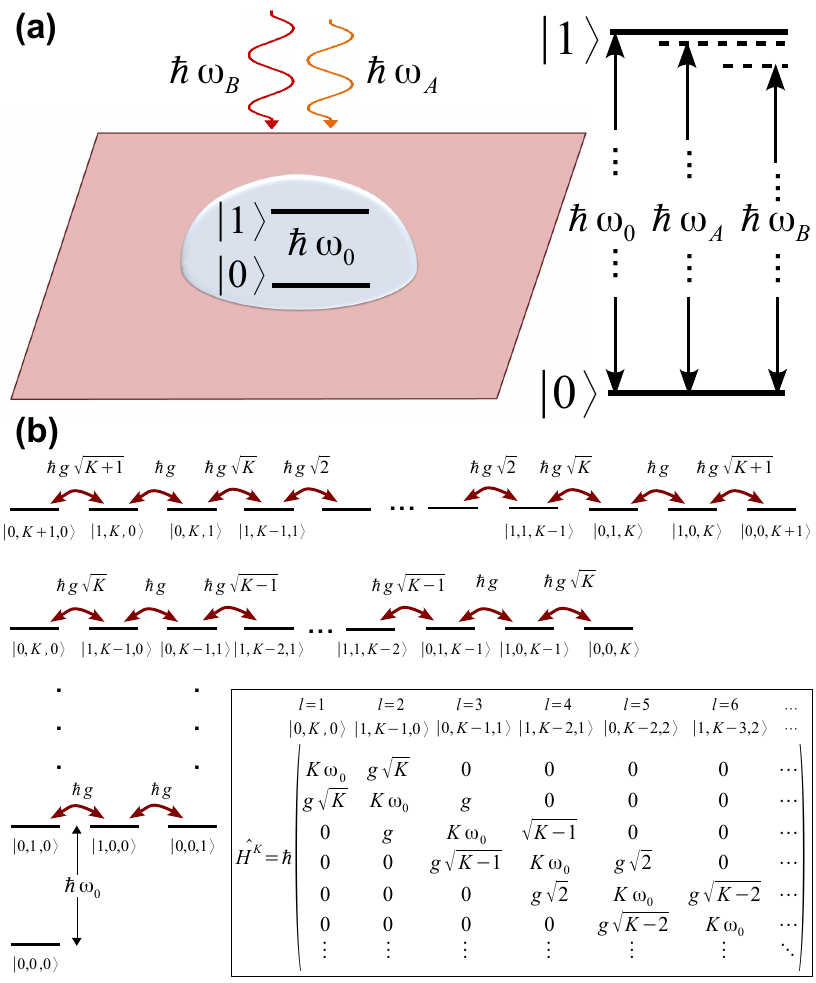}}
{\footnotesize \caption{(a) Schematics of the system and involved photon energies.  (b) Energy level structure in different manifolds before coupling (the ``hopping terms" are indicated), for the resonantly monochromatic double dressing case. The inset shows the top-left corner of the Hamiltonian for a manifold with K$>$3, where the corresponding basis element and their labels are indicated.}}
\label{fig1} 
\end{figure}

From this point we focus on the resonant monochromatic double dressing case, in which $\omega_B = \omega_A = \omega_0$. 

Given this condition, eq. (\ref{eq-1}) for each subspace becomes a tridiagonal matrix with constant diagonal term $K\hbar\omega_0$ (a manifold with degree of degeneracy $2K+1$, if the $g$ coupling is turned off). This resembles a closest-neighbor tight-binding Hamiltonian with position depending hopping \cite{tbpositiondepending}, whose diagonalization straightaway renders formation of energy bands. 

Therefore, analogously to the pass from completely discretized atomic states to band structures in crystals; it is natural to expect a drastic change in the LDOS, evolving from fully discretized levels to quasi-continuous photonic subbands.   
  
Figure 1(b) depicts the manifolds formed due to the presence of the two lasers. The couplings between levels are presented in a analogous way to the hopping terms in a first-neighbor model for one dimensional atomic chains.  

Diagonalization of this Hamiltonian [see inset in figure 1(b)] produces $2K+1$ eigenenergies ($E_l^K = \hbar g \varepsilon_l^K$) which are distributed symmetrically around $0$, along the interval $-\hbar g \sqrt{2 K} \leq E_l^K \leq \hbar g \sqrt{2 K}$.

Figure 2(a) shows a graphical representation of the matrix transformation $P$ which diagonalizes the Hamiltonian of eq. (1) in the resonant monochromatic double dressing case, while figure 2(b) presents the LDOS as function of the energy and number of photons in the laser B (for a given $K=200$). They clearly illustrate how the band structure appears as a direct consequence of the presence of laser B, so that the well defined side peaks (Mollow triplet \cite{mollow,nat-MT-1,nat-MT-2,nat-nano-ustc}) turn into energy sidebands as the number of photons in that laser increases (upper and lower subbands, respect to the initially degenerated eigenenergy $E=0$). 

\begin{figure}[t]
\scalebox{1}{\includegraphics{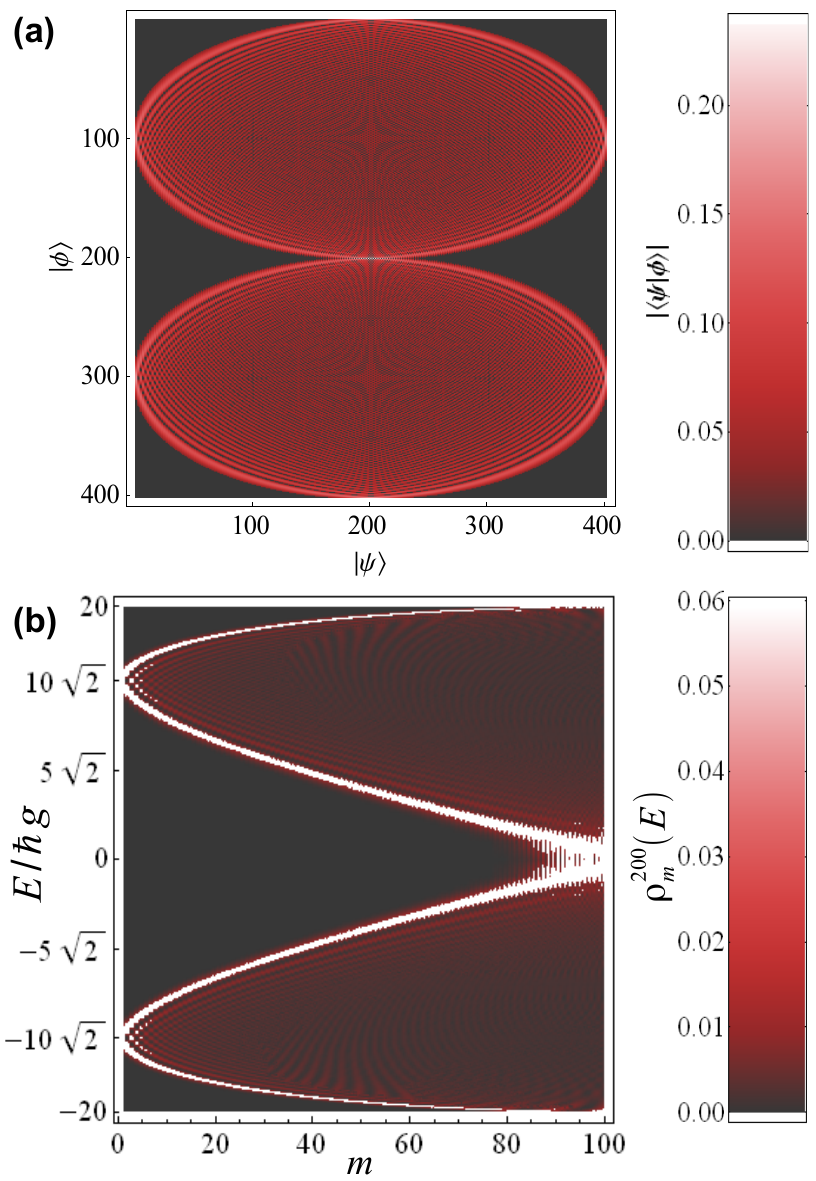}}
{\footnotesize \caption{(a) Array-plot of the unitary transformation $P$ than diagonalizes the Hamiltonian $\hat{H}$. (b) LDOS as function of the energy and number of photons in the laser B ($m$). ($K=200$. Energy has been offset, so that $K\hbar \omega_0=0$).}}
\label{fig2} 
\end{figure} 

In the purely radiative decay limit, for an exciton lifetime $\tau $, the spectral linewidths are respectively; $\Gamma_c = \frac{1}{2 \tau}$ for a transition between the upper-upper and lower-lower subbands (central peak), and $\Gamma_s =\frac{3}{4\tau}$ for a transition between the upper-lower and lower-upper subbands (side peaks) \cite{qopticsbook,mollow}.

\section{4. Analytical approximation}

For the sake of physical insight and simplicity in calculations, an analytical approximation to obtain the transformation coefficients from well known functions can be pursued.

To do this, for the $K$-th manifold we change the basis to symmetric and antisymmetric combinations of the elements of the original one; i.e. $\mid \tilde{\phi}_{m,\pm}^{K} \rangle \equiv \frac{1}{\sqrt{2}} (\mid 0,K-m,m \rangle \pm \mid ~ 1,K-m-1,m \rangle)$. Because the manifold has odd number of states, the last one in the original basis ($\mid 0,0,K\rangle$) is intentionally left unpaired for ordering the new basis in reference to it. 

In this new basis, the dot-field interaction matrix element $\langle \tilde{\phi}_{m,\pm}^K \mid \hat{H}^C \mid \tilde{\phi}_{m',\pm}^K \rangle \equiv \langle \tilde{\phi}_{m,\pm}^K \mid \hat{\sigma}_+ (\hat{a} + \hat{b}) + \hat{\sigma}_- (\hat{a}^\dag + \hat{b}^\dag) \mid \tilde{\phi}_{m',\pm}^K \rangle$, can be further evaluated to obtain

\begin{eqnarray}
\label{eq-10}
\langle \tilde{\phi}_{m,\pm}^K \mid\hat{H}^C \mid \tilde{\phi}_{m',\pm}^K \rangle &=& \pm \frac{1}{2} (2\delta_{m,m'} \sqrt{K-m'} \nonumber \\ &\hspace*{1ex}& + \delta_{m,m'-1} \sqrt{m'} + \delta_{m,m'+1} \sqrt{m'+1}) \nonumber \\
\langle \tilde{\phi}_{m,\pm}^K \mid \hat{H}^C \mid \tilde{\phi}_{m',\mp}^K \rangle &=& \pm \frac{1}{2} (\delta_{m,m'-1} \sqrt{m'} \nonumber \\ 
&\hspace*{1ex}& - \delta_{m,m'+1} \sqrt{m'+1}) \hspace*{1ex} .
\end{eqnarray}

If the new basis is ordered in the form \newline $\{ \mid \tilde{\phi}_{0,+}^{K} \rangle,\mid \tilde{\phi}_{1,+}^{K} \rangle,\mid \tilde{\phi}_{2,+}^{K} \rangle,...,\mid \tilde{\phi}_{K-1,+}^{K} \rangle,\mid 0,0,K \rangle,\mid  \tilde{\phi}_{K-1,-}^{K} \rangle,\mid \tilde{\phi}_{K-2,-}^{K} \rangle,...,\mid \tilde{\phi}_{1,-}^{K} \rangle,\mid \tilde{\phi}_{0,-}^{K} \rangle \} $. Therefore, the Hamiltonian takes the explicit form 


\begin{widetext}

\begin{equation}
\label{eq-11}
{\tiny \hat{H}^C  = \frac{\hbar g}{2} \left(
\begin{array}{ccccccccccccc}
2\sqrt{K} & 1 & 0 & \cdots & 0 & 0 & 0 & 0 & 0 & \cdots & 0 & 1 & 0 \\
1 & 2\sqrt{K-1} & \sqrt{2} & \cdots & 0 & 0 & 0 & 0 & 0 & \cdots & 2 & 0 & -1 \\
0 & \sqrt{2} & 2\sqrt{K-2} & \cdots & 0 & 0 & 0 & 0 & 0 & \cdots & 0 & -2 & 0 \\
\vdots & \vdots & \vdots & \ddots & \vdots & \vdots & \vdots & \vdots & \vdots & \iddots & \vdots & \vdots & \vdots \\
0 & 0 & 0 & \cdots & 2\sqrt{2} & \sqrt{K-1} & 0 & \sqrt{K-1} & 0 & \cdots & 0 & 0 & 0 \\
0 & 0 & 0 & \cdots & \sqrt{K-1} & 2 & \sqrt{K} & 0 & -\sqrt{K-1} & \cdots & 0 & 0 & 0 \\
0 & 0 & 0 & \cdots & 0 & \sqrt{K} & 0 & -\sqrt{K} & 0 & \cdots & 0 & 0 & 0 \\
0 & 0 & 0 & \cdots & \sqrt{K-1} & 0 & -\sqrt{K} & -2 & -\sqrt{K-1} & \cdots & 0 & 0 & 0 \\
0 & 0 & 0 & \cdots & 0 & -\sqrt{K-1} & 0 & -\sqrt{K-1} & -2\sqrt{2} & \cdots & 0 & 0 & 0 \\
\vdots & \vdots & \vdots & \iddots & \vdots & \vdots & \vdots & \vdots & \vdots & \ddots & \vdots & \vdots & \vdots \\
0 & 2 & 0 & \cdots & 0 & 0 & 0 & 0 & 0 & \cdots & -2\sqrt{K-2} & -\sqrt{2} & 0 \\
1 & 0 & -2 & \cdots & 0 & 0 & 0 & 0 & 0 & \cdots & -\sqrt{2} & -2\sqrt{K-1} & 1 \\
0 &- 1 & 0 & \cdots & 0 & 0 & 0 & 0 & 0 & \cdots & 0 & -1 & -2\sqrt{K} 
\end{array}
\right) \hspace*{1ex}.
}
\end{equation}

\end{widetext}

In the above matrix, the effects of the laser A on the QD are contained in the diagonal, where energy splittings of magnitude $\Delta E = 2 \hbar g \sqrt{n}$ are observed between the symmetric and antisymmetric combinations $\mid \hat{\phi}_{m,+}^K \rangle$ and $\mid \hat{\phi}_{m,-}^K \rangle$. 

Taking aside the diagonal part $\hat{H}^D$, the diagonalization of the remaining matrix ($\hat{H}^{ND}$) delivers the energy modifications caused by the laser B on the dressed states of the coupled system QD-laser A. Those eigenvalues and eigenvectors define the energy spectrum and LDOS of the double driven QD, symmetrically distributed around $\hbar g \sqrt{N}$ and $-\hbar g \sqrt{N} $ (upper and lower energy subbands). 

Matrix $\hat{H}^{ND}$ can be seen as composed by four equally sized blocks, plus the row and column corresponding to the state $\mid 0,0,K \rangle$. The off-diagonal terms in the top-right and bottom-left blocks (off-diagonal blocks, responsible of the mixing between the diagonal ones), increase in absolute value as their position are closer to the matrix center, i.e. $m \sim K$.

Hence, for the system in the limit $K\rightarrow\infty$ and $M/K \rightarrow 0$ (in which laser A is much more intense than laser B and then mixing effects between blocks are negligible); the eigenvectors from diagonalization of $\hat{H}^{ND}$ are a superposition of the new basis according  to

\begin{equation}
\label{eq-12}
\mid \tilde{\psi}_{l,\pm}^K \rangle = \sum_{j=1}^{K} \tilde{c}_{j,l,\pm}^{K} \mid \tilde{\phi}_{j,\pm}^{K} \rangle \hspace*{1ex} ,
\end{equation}

where the coefficients $\tilde{c}_{j,l,\pm}^{K} \equiv  \langle \tilde{\phi}_{j,\pm}^{K} \mid \tilde{\psi}_{l,\pm}^K \rangle$ are to be associated to a set of orthonormal functions.

Since the off-diagonal elements in the top-left [bottom-right] block, located exclusively right above and below the matrix diagonal, have the same structure as those of the operator $(\hat{b} + \hat{b}^\dag)\equiv \sqrt{2} \hat{x}$ [$-(\hat{b} + \hat{b}^\dag)\equiv -\sqrt{2} \hat{x}$] written in the eigenbasis of the number of photons operator ($\hat{m}=\hat{b}^\dag \hat{b}$) \cite{quantumbook};
in this limit the matrix elements of the transformation that diagonalizes $\hat{H}^{ND}$ can be obtained in good approximation from the harmonic quantum oscillator eigenfunctions $\Phi_j (z)$ \cite{atomicsimilar3,qopticsbook,quantumbook}. This is

\begin{equation}
\label{eq-13}
\tilde{c}_{j,l,\pm}^{K} \rightarrow \tilde{c}_{j,\varepsilon,\pm} = \langle j \mid \pm \varepsilon \rangle \equiv \Phi_j (\frac{\pm \varepsilon}{\sqrt{2}}) \hspace*{1ex},
\end{equation}

 with $\varepsilon\equiv\sqrt{2}x$ ($-\infty < \varepsilon < \infty$), a dimensionless continuum parameter which multiplied by $\hbar g$ becomes the eigenenergy corresponding to the eigenvector $\mid \sqrt{2} x \rangle$. The functions are explicitly 

\begin{eqnarray}
\label{eq-14}
\Phi_j (z) = \left( \frac{1}{\sqrt{\pi} 2^j j!}\right)^{1/2} \exp \left(- \frac{z^2}{2} \right) H_j(z) \hspace*{1ex} ,
\end{eqnarray}  

where $H_j(x)$ is the $j$-th order Hermite polynomial \cite{prb-hanz,quantumbook}.  

For the relevant transitions between each of the two subbands in the $K+1$ and $K$ manifolds, there are four possibilities: transitions between the upper (lower) subbands, alongside with transitions from the upper $K+1$ (lower $K+1$) subband to the lower $K$  (upper $K$) one. The dipole transition matrix elements become 
$ \mid \langle \tilde{\psi}_{F,\pm}^{K} \mid \hat{\sigma}_- \mid \tilde{\psi}_{I,\pm}^{K+1} \rangle \hspace*{1ex} \mid^2 = \pm \frac{1}{2} \delta(\varepsilon_F - \varepsilon_I)$ and 
$ \mid \langle \tilde{\psi}_{F,\mp}^{K} \mid \hat{\sigma}_- \mid \tilde{\psi}_{I,\pm}^{K+1} \rangle \hspace*{1ex} \mid^2 = \pm \frac{1}{2} \delta(\varepsilon_I + \varepsilon_F \pm 2 \sqrt{N})$, respectively; 
where the transition energy has been also shifted to make $\hbar \omega_0 = 0$.

On the other side, the normalized LDOS for the reference number of photons $M$ as function of the energy of the transition final state ($E_F = \hbar g \varepsilon_F$), is now given by 

\begin{equation}
\label{eq-15}
\rho_{M}(E_F)=(\frac{\sqrt{2}}{\hbar g}) \mid \Phi_M (\frac{\sqrt{2} E_F}{\hbar g}) \mid ^2 \hspace*{1ex}.
\end{equation}

Figure 3 shows the LDOS as function of the emission frequency normalized to the coupling $g$, for different numbers of photons in the laser B. It is worth noting how the LDOS has absolute maxima at values close to $\omega=\pm{\sqrt{M}}g$, respectively \cite{correspondence}. For frequencies more separated from the reference ($\omega/g = 0$), it decays rapidly to zero; setting an approximate width for the subbands of two-times the Rabi splitting associated to the laser B.   

\begin{figure}[h]
\scalebox{1}{\includegraphics{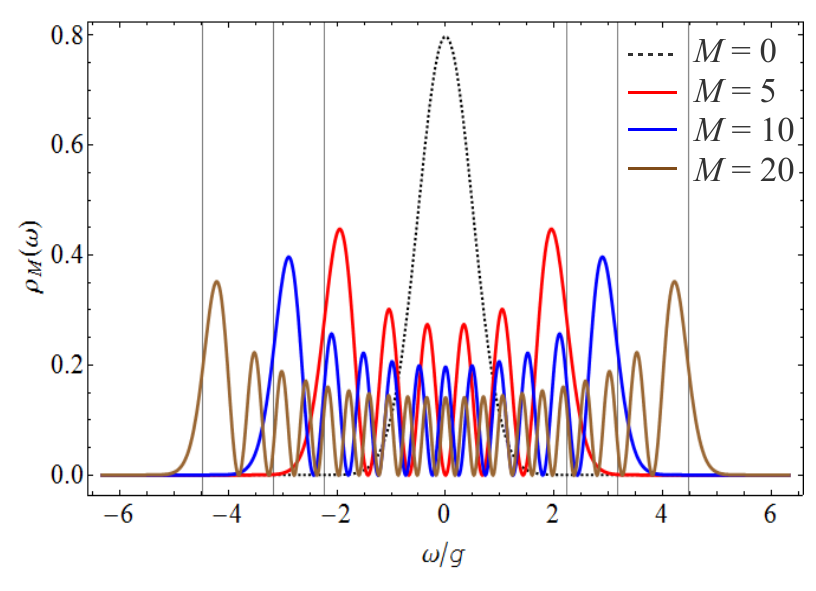}}
 \caption{LDOS for different numbers of photons in the laser B. The thin vertical lines indicate the corresponding values $\omega=\pm\sqrt{M}g$. The dashed black line shows the case in which only laser A is applied ($M=0$).}
\label{fig3} 
\end{figure}  

Under these considerations, the FGR yields for the emission spectrum 

\begin{eqnarray}
\label{eq-17}
I(\omega) &\propto& \int_{-\infty}^{\infty} \int_{-\infty}^{\infty} d E_I \hspace*{1ex} d E_F \hspace*{1ex} [ \delta(E_I + E_F + 2 \hbar g \sqrt{N}) + 2 \delta(E_I - E_F)\nonumber \\
&\hspace*{1ex}&  + \delta(E_I + E_F - 2 \hbar g \sqrt{N}) ]  \hspace*{1ex} \rho_{M}(E_F) \hspace*{1ex} L(\hbar\omega + E_F - E_I,\Gamma_{I,F}) \hspace*{1ex} , 
\end{eqnarray}

which after insertion of eq. (\ref{eq-15}) and integration over $E_I$ turns in

\begin{eqnarray}
\label{app-2}
I(\omega) &\propto& \int_{-\infty}^{\infty} d E_F \hspace*{1ex} [ \frac{\hbar \Gamma_s}{(\hbar \omega + 2E_F + 2\hbar g \sqrt{N})^2 + (\hbar \Gamma_s)^2} + \frac{2 \hbar \Gamma_c}{(\hbar \omega)^2 + (\hbar \Gamma_c)^2} \nonumber \\
&\hspace*{1ex}& + \frac{\hbar \Gamma_s}{(\hbar \omega + 2E_F - 2\hbar g \sqrt{N})^2 + (\hbar \Gamma_s)^2} ]  \hspace*{1ex} \mid \Phi_M(\frac{\sqrt{2} E_F}{\hbar g}) \mid^2  \hspace*{1ex} .
\end{eqnarray}

\begin{figure}[h]
\scalebox{1}{\includegraphics{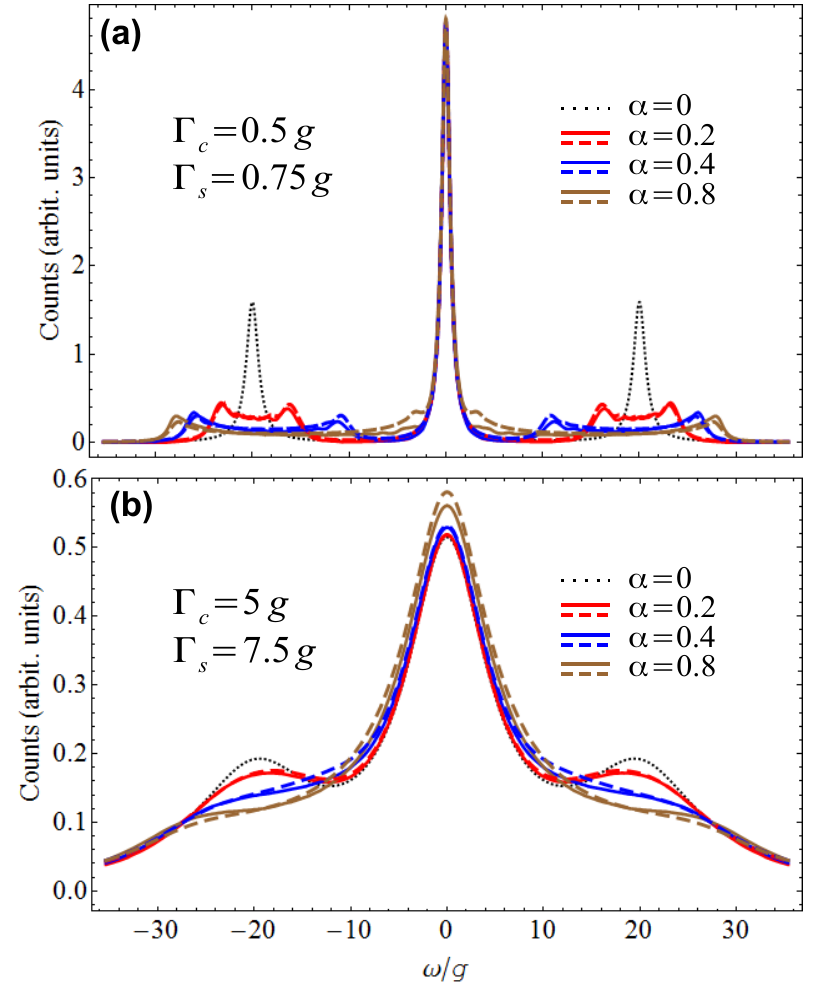}}
 \caption{Calculated resonance fluorescence spectra for different laser intensity ratios; given a fixed $K=100$. The solid lines are obtained from eq. (\ref{app-1}) while the dashed lines are obtained from eq. (\ref{app-2}). The dotted black line shows the well known Mollow triplet case, when just laser A is applied. (a) $\Gamma_c = 0.5 g$ and $\Gamma_s = 0.75 g$, (b) $\Gamma_c = 5 g$ and $\Gamma_s = 7.5 g$. }
\label{fig4} 
\end{figure}

In figure 4 the resonance fluorescence spectra calculated for different ratios between the numbers of photons in lasers A and B ($\alpha^2\equiv\frac{m}{n}$), are shown for a fixed total number of photons $K=100$. They are obtained by using both, the numerical quasi-exact approach of eq. (\ref{app-1}) and the analytical approximation of eq. (\ref{app-2}).
In the upper frame, for a $\Gamma_c$ close to the value of the coupling constant $g$ (i.e. the exciton lifetime is large enough respect to the Rabi oscillation periods defined by the laser intensities); it is clearly appreciated how the Mollow side peaks of the single excitation laser case evolve into sidebands, reflecting the energy band formation under the double dot-field coupling. Markedly the subband width increases with the number of photons in the laser B.

In other words, due to the second driving field, well defined peaks spread into optically active regions; then exhibiting similarities with LDOS proper of higher dimensionality systems \cite{biggerdimension1,biggerdimension2,biggerdimension3}.     

When $\Gamma_c$ is substantially larger than $g$, the effects of the second laser become less noticeable and useful. This because in the time dominion, the short exciton lifetime inhibits coherent Rabi oscillations making almost irrelevant the presence of laser B. This sensitivity of the system to the ratio between $g$ and $\Gamma_c$ provides a way to estimate the order of magnitude of the dot-field coupling.       

Figures 4(a) and 4(b), evidence that the analytical approximation behind eq. (\ref{app-2}),  as long as the value of $\alpha$ is not close to one, works ostensibly well even for a moderate number of photons. The main discrepancy is found around the subband edges, where the exact calculation predicts slight asymmetry due to the finite nature of the Gilbert subspaces. Whereas such an asymmetry is tenuous, actually it has been experimentally observed \cite{parallel}.       

\section{5. Conclusions}

In conclusion, a theoretical approach to model and simulate doubly driven artificial atoms was implemented.  

As a main result, tailored manipulation of the optical density of states in semiconductor quantum dots, has been shown for the case of  monochromatic double dressing. It was described how by coupling a nanostructured qubit simultaneously to two distinguishable lasers whose frequencies match the exciton transition, a discrete eigenstate turns into an energy subband in a process closely analogous to band formation in solid state physics. 

Such strong changes in the local density of optical states, controllable through the ratio between the laser intensities; open new possibilities for on-demand photon emission from artificial atoms.

The presented results are in remarkable qualitative and quantitative agreement with experimental measurements, as presented in Ref. \cite{parallel}. 


The author thanks the Chinese Academy of Sciences for financial support through the ``Fellowship for Young International Scientist", Grant No. 2011Y1JB03. Valuable discussion with C.Y. Lu's group is acknowledged.


\begin{thebibliography}{99}
\bibitem{correspondence} This is related to the correspondence principle for the density of probability in the quantum harmonic oscillator \cite{quantumbook}.
\end{thebibliography}

\end{document}